\RequirePackage[l2tabu,orthodox]{nag}
\RequirePackage{fix-cm}
\documentclass[letterpaper,10pt,journal,compsoc, final]{IEEEtran}
\usepackage[utf8]{inputenc}
\usepackage{textcomp}
\ifCLASSOPTIONcompsoc
\usepackage[nocompress]{cite}
\else
\usepackage{cite}
\fi
\ifCLASSINFOpdf
\usepackage[pdftex]{graphicx}
\graphicspath{{../images/}}
\DeclareGraphicsExtensions{.png, .pdf}
\else
\usepackage[dvips]{graphicx}
\graphicspath{{../eps/}}
\DeclareGraphicsExtensions{.eps}
\fi
\usepackage{amsmath}
\usepackage{array}
\usepackage{tabularx}
\usepackage{makecell}
\usepackage{dirtytalk}
\usepackage{caption}
\usepackage{booktabs, multicol, multirow}
\usepackage{rotating}

\usepackage{subcaption}
\ifCLASSOPTIONcaptionsoff

\usepackage[nomarkers]{endfloat}
\let\MYoriglatexcaption\caption
\renewcommand{\caption}[2][\relax]{\MYoriglatexcaption[#2]{#2}}
\fi
\usepackage[ruled,vlined, longend]{algorithm2e}

\usepackage{blindtext}
\usepackage{enumitem}
\usepackage [english]{babel}
\usepackage [autostyle, english = american]{csquotes}

\usepackage{eulervm}   
\usepackage[activate={true,nocompatibility},final,tracking=true,kerning=true,spacing=true,factor=1100,stretch=10,shrink=10]{microtype}

\usepackage[usenames,dvipsnames]{xcolor}
\definecolor{LinkColor}{RGB}{216,0,0}

\usepackage{glossaries}

\usepackage{indentfirst}
\glsdisablehyper
\usepackage[hang,flushmargin]{footmisc}
\usepackage{enumitem}
\newlist{steps}{enumerate}{1}
\setlist[steps, 1]{label = Step \arabic*:}

\usepackage[pdftex, breaklinks]{hyperref}
\hypersetup{ 
	bookmarksnumbered=true,
	colorlinks=true,
	linkbordercolor =White, 
	urlcolor=Gray, 
	filecolor=LinkColor,
	citebordercolor=LinkColor,
	citecolor=LinkColor,	linkcolor=LinkColor,  
	pdftitle=\@title
}
\usepackage[nameinlink,capitalise]{cleveref}

\usepackage{balance}

\newlength\mylen

\definecolor{myGray}{HTML}{555555}
\definecolor{algoColorComment}{named}{myGray}

\definecolor{arsenic}{rgb}{0.23, 0.27, 0.29}

\def \datasetUrl {https://www.kaggle.com/qiriro/ieee-tac}

\definecolor{aurometalsaurus}{rgb}{0.43, 0.5, 0.5}
\newcommand{\myFooterTex}[1]{\footnote{\color{Gray}{#1}}}

\begin{document}

\title{The Effect of Person-Specific Biometrics in Improving Generic Stress Predictive Models}
\author{
	\IEEEauthorblockN{Kizito~Nkurikiyeyezu\thanks{corresponding author: e-mail \textemdash kizito@wil-aoyama.jp}, Anna~Yokokubo, and Guillaume~Lopez}
	\IEEEauthorblockA{\\Wearable Environment and Information Systems Lab\\
		Aoyama Gakuin University\\
		5-10-1 Fuchinobe Chuo-ku, Sagamihara-shi, Kanagawa-ken 252-5258, Japan}
	 \vspace*{-0.5cm}
}
\markboth{Journal of Sensors and Materials (In press, Spring 2020)}%
{Nkurikiyeyezu \MakeLowercase{\textit{et al.}}: The effect of  person-specific biometrics}
\maketitle

\begin{abstract}
Because stress is subjective and is expressed differently from one person to another, generic stress prediction models (i.e., models that predict the stress of any person) perform crudely. Only person-specific ones (i.e., models that predict the stress of a preordained person) yield reliable predictions, but they are not adaptable and costly to deploy in real-world environments. For illustration, in an office environment, a stress monitoring system that uses person-specific models would require collecting new data and training a new model for every employee. Moreover, once deployed, the models would deteriorate and need expensive periodic upgrades because stress is dynamic and depends on unforeseeable factors. We propose a simple, yet practical and cost-effective calibration technique that derives an accurate and personalized stress prediction model from physiological samples collected from a large population. We validate our approach on two stress datasets. The results show that our technique performs much better than a generic model. For instance, a generic model achieved only a $42.5\%\pm 19.9\%$ accuracy. However, with only 100 calibration samples, we raised its accuracy to $95.2\%\pm0.5\%$. We also propose a blueprint for a stress monitoring system based on our strategy, and we debate its merits and limitation. Finally, we made public our source code and the relevant datasets to allow other researchers to replicate our findings. 
\end{abstract}
\ifCLASSOPTIONpeerreview
 \begin{center} \bfseries EDICS Category: 3-BBND \end{center}
 \fi
\IEEEpeerreviewmaketitle

\section{INTRODUCTION}\label{sec:intro}
\par
\IEEEPARstart {O}{cupational} stress is well-researched \cite{Carneiro2019} \cite{Can2019a} \cite{Schmidt2018}\cite{Carneiro2017}\cite{Alberdi2016}, though not least due to its pernicious effect on people's health but also due to the economic benefits of keeping in check the stress level of employees. Admittedly, although a small amount of stress is benign and even auspicious because it provides the necessary gumption to survive the tribulations of the modern workplace \cite{Kirby2013} \cite{Dhabhar2012}, chronic stress (i.e., enduring stress) has detrimental repercussions. Physiological and psychological disorders \cite{Tennant2001} \cite{Colligan2005a}, job-related tensions \cite{Ganster2013}, and general deterioration of health are just a few examples of its adverse outcomes. Furthermore, stress is liable for significant economic losses because stressed-out workers have suboptimal productivity, are prone to higher job absenteeism and presenteeism,  and are disproportionately predisposed to sickness  \cite{EU-OSHA2017, Colligan2005a}.
\par
Consequently, the importance of overcoming stress at work is primordial to the well-being of the workers and the bottom line of any business. Nevertheless, at the moment, there exist no mainstream real-world stress monitoring system \cite{Peake2018}. The most reliable stress monitoring strategies rely on directly measuring the level of the stress-inducing hormones (e.g., salivary and cortisol concentration in sweat \cite{Marques2010} \cite{Hellhammer1994}) and on psychological evaluations performed by psychologists. However, these procedures are neither suitable nor feasible for continuously monitoring stress in the workplaces because they are obtrusiveness and are carried out sporadically. Moreover, in the case of physiological evaluations, people are reluctant to reveal their work stress honestly  \cite{Eisen2008}. Luckily, stress spawns detectable physiological, psychological, and behavioral changes that can be used for automatic stress recognition \cite{Carneiro2019} \cite{Alberdi2016}. For example, acute stress decreases a person's Heart Rate Variability (HRV) and his parasympathetic activation \cite{JARVELIN-PASANEN2018}. Besides, there is plentiful research that shows that it is plausible to indirectly monitor stress using physiological signals such as the Electrodermal Activity (EDA) \cite{Adams2014}, the HRV \cite{Melillo2011}\cite{Cinaz2013}, the Electroencephalogram (EEG)\cite{Rahnuma2011}, and the Electromyography (EMG)\cite{Wei2013}.
\par
Although there is a surfeit of publications \cite{Carneiro2019} \cite{Schmidt2018} \cite{Poria2017} \cite{Alberdi2016} on automatic stress prediction, at the moment, aside from a few niche and non-scientifically proven consumer products, there exist no effective system that automatically and unobtrusively monitor people's stress in real-world environments \cite{Peake2018}. On the one hand, some of the proposed approaches (e.g., EEG based stress monitoring) are outright impractical because they are too obtrusive.  On the other hand, the most precise approaches (e.g., \cite{Koldijk2018}, \cite{Poria2017} and \cite{Mozos2017}) predict stress using a fusion of multiple sensors data (e.g., audio, video, computer logging, posture, facial expression, and physiological features). These methods, however, raises technical, privacy and security challenges (e.g., the implication of user's computer keystrokes logging, video recording, and speech recording), and,  are therefore inconvenient to deploy in the real-world settings because of company-wide computer security policies or due to international workplace privacy regulations. Finally, the most practical and unobtrusive stress monitoring methods (e.g., \cite{Gjoreski2016}\cite{Kocielnik2013} \cite{Zhai2006}\cite{Healey2005})\textemdash which are mostly based on physiological signal that are recordable on people's wrist (e.g., Photoplethysmography (PPG) and EDA) \textemdash are not yet mainstream to the general consumers despite their potential economic and health benefit. The lack of a viable stress monitoring products, despite the extensive research on occupational stress, the availability of enabling technology (e.g., smartphones with on-wrist HRV and EDA sensors)  and despite the immense economical and health benefits such products would bring, begs the question of why this is the case. 
\par 
A recent review article on affect and stress recognition \cite{Schmidt2018} scrutinized the published literature and noted the striking discrepancy between the accuracy of person-specific stress prediction Machine Learning (ML) models (i.e., ML models that predict the stress of a specific person) and person-independent ML  models (i.e., generic ML models that predict the stress of a any person). The article underscores that person-specific ML models (e.g.,\cite{Koldijk2018}, \cite{Nakashima2016}, \cite{Picard2001},\cite{Healey2005},\cite{Haag2010},\cite{Melillo2011},\cite{Valenza2014a} and \cite {Alberdi2018}) achieved an excellent prediction accuracy. Nevertheless, their predictions are person-specific\textemdash that is, the ML models would not generalize well in predicting stress of yet unseen people; therefore, cannot be used in creating mass-market stress monitoring products. On the contrary, the pragmatic person-independent solutions (e.g., \cite{Schmidt2018a},\cite {Koldijk2018},\cite{Zenonos2016},\cite{Nakashima2016}, \cite{Gjoreski2016}, \cite{Kolodyazhniy2011}, and \cite{Andre2008}) generally have a much lower stress prediction accuracy; accordingly, they are equally a poor choice for creating mass-market stress monitoring products.  For example, \cite{Andre2008} achieved a 95.0\% emotion recognition accuracy using person-specific ML models; however, the same approach resulted in a mere 70\%  accuracy when applied to a person-independent classification model. In a like manner, the authors in \cite{Nakashima2016} conducted experiments to monitor stress in daily work and found that ML models that use people's physiology to predict stress are highly person-dependent. Their person-specific ML models achieved a 97\% accuracy but the generic ones dwindled to a mere 42\% accuracy. Their results resemble that in \cite{Koldijk2018}, which achieved a 90.0\% accuracy when using a person-specific stress classification models. However, when applied the same approach to predict the stress of new subjects, its performance ebbed to a meager 58.8$\pm$11.6\% accuracy. 
\par \label{sec:intro:underperformance}
These mediocre outcomes are expected. For example, the authors in \cite{Lamichhane2017} argued that, when people's physiological differences are not accounted for, the ML stress prediction models performed no better than a model with no learning capability. First stress is intrinsically idiosyncratic and depends on a person's uniqueness (e.g., his genetics) and his coping ability \cite{Kogler2015}. Second, there is incontrovertible evidence that there exist gender differences in how people respond to stress \cite{Wang2007a} and that men and women have a different feeling about stress because women tend to express a higher level of stress on self-report questionnaires \cite{Liapis2015} \cite{Matud2004}. Third, a stressor that produces stress in one person will not necessarily trigger the same stress response in a different person \cite{Hernandez2011}\cite{Childs2014}\cite{Johnstone2015} \cite{Sapolsky1994}. Finally, for the same person, there exist significant day-to-day variability in the cortisol awakening response, which may affect how that person responds to stress \cite{Almeida2009}. As a result, a practical stress monitoring scheme needs to take into account inter-individual and intra-individual differences, people's gender, the temporal variability of human stress and many other factors that influences how humans react to stress. The state of the art stress monitoring strategies (e.g.,  \cite{Attaran2018}) use person-specific ML models. Unfortunately, this method is not realistic for creating a real-world product. A stress monitoring system that uses this approach would be costly (e.g., collecting and training ML stress prediction models for every user of the system) and would require expensive recurrent updates because stress is innately dynamic. 
\par 
The recent research has proposed diverse methods to improve the performance of the generic stress prediction models. The most straightforward methods use normalization techniques (e.g., range normalization, standardization, baseline comparison, and Box-Cox transformation) to reduce the impact of inter-individual variability while preserving the differences between the stress classes \cite{Lamichhane2017}\cite{Aigrain2016}. The normalization improves the performance of the generic model but always underperforms compared to the person-specific ones. Furthermore, as \cite[Chap.~5]{Aigrain2016} noted, the normalization process is multifaceted and depends on trial and error methods. An alternative strategy is to predict stress based on clusters of similar users \cite{Koldijk2018}\cite{Xu2015}\cite{Ramos2014}. These techniques are important contributions to producing an effective stress monitoring system. However, they also perform inadequately compared to person-specific models. Moreover, these methods would likely prove too complex to use in real-world settings because they are sensitive to the number of clusters \cite{Xu2015} and, given that many factors influence a person's stress \cite{Schneiderman2008}, it is not clear what are the criteria for similarity to create the cluster similarities.
\par 
In this paper, we propose a hybrid and cheaper to deploy stress prediction method that incorporates tiny person-specific physiological calibration samples into a much larger generic sample collected from a large group of people. The proposed method hinges on the premise that all humans share a hormonal response to stress \cite{Charmandari2005}, but that a person's unique factors such as gender \cite{Wang2007a}, genetics\cite{Wust2004}, personality \cite{Childs2014}, weight \cite{Jayasinghe2014}, and his/her coping ability \cite{Kogler2015} differentiate how the person reacts to stress. Hence, we hypothesize that it could be possible to reuse generic samples collected from many people as a starting point for creating a personalized and more effective model. To confirm these assumptions, we tested this strategy on two major stress datasets.  Our results show a substantial improvement in the stress prediction models' performance even when we used only 100 calibration samples. In summary, in this paper: 
\begin{enumerate}[label=(\roman*)]	
	\setlength\itemsep{1em}
	\item For each subject in the datasets, we train and validate \textit{n} person-specific regression and classification stress prediction models using a 10-fold cross-validation approach. The result shows that, for all subjects, the classification models achieved a greater than 95\% classification accuracy and that the regression models had a near-zero mean absolute error (MAE).  
	\item We used a Leave-One-Subject-Out Cross-Validation (LOSO-CV) to assess the performance of generic stress prediction models.  All models performed poorly (e.g., $42.5\%\pm 19.9\%$  accuracy, $14.0\pm7.9$ MAE, on one dataset) compared to person-specific models and that there was a wide performance variation between the subjects. 
	\item We devise a hybrid technique that derives a personalized person-specific-like stress prediction model from samples collected from a large population and discussed how it could be used to develop a real-world continuous stress monitoring system in, e.g., intelligent buildings.  
\end{enumerate}
\section{METHODS}
\begin{table*}[htbp]
	\centering
	
	\caption{Selected heart rate variability  (HRV) and electrodermal activity (EDA) features}
	\vspace{-2mm}
	\begin{tabular}{@{}rllr@{}}
		\toprule
		\multicolumn{1}{c}{\multirow{9}[4]{*}{\begin{sideways}HRV Features\end{sideways}}} & \multicolumn{1}{p{6.335em}}{Time domain} & Mean, median, standard deviation, skewness and kurtosis of all RR intervals &  \\
		& \multicolumn{1}{p{6.335em}}{RMSSD} & Root mean square of the successive differences &  \\
		& \multicolumn{1}{p{6.335em}}{SDSD} & Standard deviation of all interval of differences between adjacent RR intervals &  \\
		& \multicolumn{1}{p{6.335em}}{SDRR\_RMSSD} & Ratio of SDRR over RMSSD &  \\
		& \multicolumn{1}{p{6.335em}}{pNNx} & Percentage of number of adjacent RR intervals differing by more than 25 and 50 ms &  ref. \cite{TaskForce1996} \\
		& \multicolumn{1}{p{6.335em}}{SD1, SD2} & Short and long-term poincare plot descriptor of the heart rate variability  & \\
		& \multicolumn{1}{p{6.335em}}{RELATIVE\_RR} & Time domain features(e.g., mean, median, SDRR, RMSSD) of the relative RR & see note a \\
		& \multicolumn{1}{p{6.335em}}{VLF, LF, HF} & Very low (VLF), Low (LF), High (HF)  frequency band in the HRV power spectrum &  \\
		& \multicolumn{1}{p{6.335em}}{LF/HF} & Ration of low (LF) and high(HF) HRV frequencies &  \\
		\cmidrule{1-3}    \multicolumn{1}{c}{\multirow{8}[2]{*}{\begin{sideways}EDA Features\end{sideways}}} & Time domain & Mean, max, min, range, kurtosis, skewness of the SCR &  \\
		& Derivatives & Mean and standard deviation of the 1st and second derivative of the SCR &  \\
		& \multicolumn{1}{p{6.335em}}{Peaks} & Mean, max, min, standard deviation of the peaks &  \\
		& \multicolumn{1}{p{6.335em}}{Onset} & Mean, max, min, standard deviation of the onsets &  ref. \cite{Zangroniz2017} \\
		& ALSC  & Arc length of the SCR &  see note b\\
		& INSC  & Inegral of the SCR & see note c \\
		& APSC  & Normalized average power of the SCR & see note d \\
		& RMSC  & Normalized room mean square of the SCR & see note e \\
		\cmidrule{1-3}                  
		\multicolumn{3}{l}{ $^a REL_{{RR}_{i}}=2\Big[\frac{RR_{i} -RR_{i-1}}{RR_{i} +RR_{i-1}}\Big], \quad i=2, ..., N \quad $ } &  \\
		\multicolumn{3}{l}{$^b
			ALSC=\sum_{n=2}^{N}\sqrt{1+\big(r[n]-r[n-1]\big)^{2}}$} &  \\
		\multicolumn{3}{l}{$^c INSC=\sum_{n=1}^{N}\big|r[n]\big|$} &  \\
		\multicolumn{3}{l}{$^d APSC=\frac{1}{N}\sum_{n=1}^{N}r[n]^{2}$} &  \\
		\multicolumn{3}{l}{$^e RMSC=\sqrt{\frac{1}{N}\sum_{n=1}^{N}r[n]^{2}}$} &  \\ 
		\bottomrule
	\end{tabular}%
	\label{tab:hrv-eda-features}%
	\vspace{-4mm}
\end{table*}%

\subsection{Stress datasets} \label{sec:dataset}
\par 
We used two stress datasets to conduct this study. The first dataset \textemdash the SWELL dataset \cite{Koldijk2014} \textemdash was collected at the Radboud University. This dataset is a result of experiments conducted on 25 subjects doing office work (for example writing reports, making presentations, reading e-mail and searching for information) who were exposed to quintessential work stressors (e.g., being unexpectedly interrupted by an urgent e-mail and pressure to complete work in a limited time). During the experiment, the researchers recorded the subjects' computer usage patterns, their facial expressions, their body postures, their electrocardiogram (ECG) signal, and their electrodermal activity (EDA) signal. The participants went through three different working conditions: 
\begin{enumerate}
	\setlength\itemsep{1em}
	\item \textit{no stress} \textemdash the participants performed the assigned tasks for a maximum of 45 minutes.  
	\item\textit{time pressure} \textemdash each participant's time to finish the task was reduced to two-thirds of the duration that he/she took in the no-stress condition.
	\item \textit{interruption} \textemdash the participants received interrupting e-mails in the middle of their assigned tasks. Some e-mails were relevant to their tasks, and the participants were requested to take specific actions. Other e-mails were immaterial, and the participants did not need to take any action. 
\end{enumerate}
At the end of each experiment condition, each participant's perceived stress was assessed using a variety of self-report questionnaires, including the NASA Task Load Index (NASA-TLX) \cite{Hart1988}. In this study, we focus on the NASA-TLX because it indicates a person's mental load based on a weighted average of multi-dimensional rating (in terms of mental demand, physical demand, temporal demand, effort, performance, and frustration) and is the standard method in assessing subjective workload.
\par 
The second dataset \textemdash the WESAD dataset \cite{Schmidt2018a}\textemdash was collected by researchers from the Robert Bosch GmbH and the University of Siegen in Germany. The dataset includes physiological (EDA, ECG, EMG, respiration signal and skin temperature) and acceleration signal that the researchers collected from 15 subjects to whom they exposed to three affective stimuli as follows:
\begin{enumerate}
	\setlength\itemsep{1em}
	\item \textit{baseline condition}\textemdash the baseline condition aimed at generating a neutral affective state onto the participants and lasted for 20 minutes.
	\item \textit{amusement condition} \textemdash the subjects watched funny video clips. Each video clip is followed by a brief (5 seconds) of neutral condition. The amusement condition lasted 392 seconds. 
	\item stress conditions \textemdash the participants were subjected to the Trier Social Stress Test (TSST) \cite{Kirschbaum1993} and asked to give a five-minutes public speech and to count down from 2023 by 17. If the subject made an error, he/she is requested to start over. 
\end{enumerate}
The amusement and the stress conditions were each followed by a meditation period to ``de-excite'' the participants back to the baseline conditions. Throughout the experiment, the participants provided five self-reports, including the Short Stress State Questionnaire (SSSQ) \cite{Helton2015} which was used to determine the type of stress (i.e., worry, engagement or distress) that was prevalent in the participants. 
\subsection{Feature extraction} \label{sec:feature-computation}
\par 
We extracted HRV and EDA features from the two datasets. We computed the HRV features according to the standards and algorithms proposed by the Task Force of the European Society \cite{TaskForce1996}. Each HRV feature (\cref{tab:hrv-eda-features}) was computed on a five-minutes moving window as follows: first, we extracted an Inter-Beat Interval (IBI) signal from the peaks of the Electrocardiogram (ECG) signal of each subject. Then, we computed each HRV index on a 5-minutes IBI array. Finally, a new IBI sample is appended to the IBI array while the oldest IBI sample is removed from the beginning of the IBI array. The new resulting IBI array is used to compute the next HRV index. We repeated this process until the end of the entire IBI array. Likewise, for the EDA signals, the raw EDA signal was first filtered by a 4Hz fourth-order Butterworth low pass filter and then smoothed with a moving average filter. Next, we computed the EDA features ( \cref{tab:hrv-eda-features}) on 10-minute moving window signal extracted from various EDA attributes of the skin conductance response (SCR). 
\par
All the resulting datasets\textemdash especially the WESAD datasets\textemdash are inherently unbalanced because their experimental protocols dictated different duration. We downsampled the datasets by randomly discarding some samples from the majority classes to make the dataset balanced; therefore, to prevent the majority classes from overshadowing the minority classes. Furthermore,  for the WESAD dataset, we altogether removed all sample corresponding to \textit{amusement condition} because it is almost as short as the sliding window we would use for computing the feature.
\subsection{Feature engineering} \label{sec:feature-engineering} 
An inspection of the histogram plots of the features computed in section \ref{sec:feature-computation} revealed that most features' data distribution is skewed. While this may not be an issue for some machine learning algorithms, in other cases, the distribution of the features is critical. For example, linear regression models expected a Gaussian distributed dataset. We mitigated this risk by applying a logarithmic transformation, a square root transformation, and a Yeo and Johnson \cite{Yeo2004} transformations to the skewed features. The application of the three transformations aimed to mutate the dataset into a new dataset that can be used with most machine learning algorithms. The logarithmic transform shrinks long heavy-tailed distribution of a feature X and bolsters its smaller values into larger ones. Therefore, it roughly transforms the data distribution into a normal distribution and reduces the effect of outliers. Likewise, we applied a square root transform on all positive feature to magnify the features' small numbers and to counterweight larger ones. However, it not possible to apply neither the logarithm transformation nor the square root transform to negative values; therefore,  we used a Yeo and Johnson (\cref{eq:yeo-johnson}) transformation to the negative skewed features.
\begin{equation}\label{eq:yeo-johnson}
y(\lambda)= 
\begin{cases} 
\frac{(y+1)^{\lambda}-1}{\lambda},& \text{when } \lambda  \neq0, \quad y\geq 0\\
log(y+1), & \text{when } \lambda  =0, \quad y\geq 0\\
\frac{(1-y)^{2-\lambda}-1}{\lambda-2},& \text{when } \lambda  \neq2, \quad y< 0\\
-log(1-y),& \text{when } \lambda =2, \quad y< 0\\
\end{cases}
\end{equation}
Additionally, as suggested in \cite{Aigrain2016}\cite{Lamichhane2017}, to minimize the influence of outliers and the inter-individual physiological variation in adapting to a stressor, we scaled the datasets by applying a scaler $S_c(X)$ to every data point $X_i$ of each feature $X$ (\cref{eq:scaler}). $S_c(X)$ removed the feature's media and uses its $25^{th}$ and $75^{th}$ quantiles to re-adjust the data points. 
\begin{equation}\label{eq:scaler}
S_c(X)=\frac{X_{i}-median(X)}{Q_{3}(X)-Q_{1}(X)}
\end{equation}
\par 
The feature engineering resulted in as much as 94 features. It is possible that some of these features have correlations with others and that some are not very relevant to the stress prediction. There might thus a need to decrease the number of the datasets' attributes \textemdash not least because this will reduce the computational requirements of the resulting predictive models \textemdash but most importantly because it could increase the models' generalization. We computed the mean decrease impurity (MDI) of each feature (\cref{gini}), i.e., the mean loss in impurity index of all tree of a random forest when that particular feature is used during tree splitting.
\begin{equation}\label{gini}
G_{k} =\sum_{k=1}^{K}p_{k}\big(1-p_{k}\big)
\end{equation}
Where K is the total number of features and $p_k$ the proportion of a single HRV feature k. We ranked all the features and heuristically selected only the features with high MDI and removed those with very small ones. \cref{tab:datasets-summary} summarizes the resulting datasets\myFooterTex{The dataset is available at \url{\datasetUrl}}.
\begin{table}[htbp]
	\centering
	\caption{Summary of the downsampled datasets}
	\resizebox{\columnwidth}{!}{%
		\begin{tabular}{@{}rrrrr@{}}
			\toprule
			& signal & \# of samples & \# of features & \# of classes \\
			\cmidrule{2-5}\multirow{2}[2]{*}{SWELL} & HRV   & 204885 & 75    & 3 \\
			& EDA   & 51741 & 46    & 3 \\
			\midrule
			\multirow{2}[2]{*}{WESAD} & HRV   & 81892 & 40    & 2 \\
			& EDA   & 20496 & 45    & 2 \\
			\bottomrule
		\end{tabular}%
		
		\label{tab:datasets-summary}%
	}
\end{table}%
\begin{table}[htbp]
	\centering
	\caption{Hyperparameters of the Random Forest models}
	\resizebox{\columnwidth}{!}{%
		\begin{tabular}{@{}lrr@{}}
			\toprule
			Hyperparameters & Classification & Regression \\
			\midrule
			number of trees  & 1000  & 1000 \\
			maximum depth of the trees & 2     & 2 \\
			best split max features & $\sqrt{\text{number of features}}$& $\frac{1}{3}(\text{number of features})$ \\
			\bottomrule
		\end{tabular}%
	}
	\label{tab:rf-parameters}%
\end{table}%
\begin{table}[htbp]
	\centering
	\caption{Hyperparameters of the ExtraTrees models}
	\resizebox{\columnwidth}{!}{%
		\begin{tabular}{@{}lrr@{}}
			\toprule
			Hyperparameters & Classification & Regression \\
			\midrule
			number of trees  & 1000   & 1000 \\
			maximum depth of the trees & 16    & 16 \\
			best split max features &$\sqrt{\text{number of features}}$ &$\frac{1}{3}(\text{number of features})$  \\
			\bottomrule
		\end{tabular}%
	}
	\label{tab:extra-trees-parameters}%
\end{table}%

\subsection{Stress prediction} \label{sec:classification} 
We developed regression stress prediction models based on each participant's self-reported stress and mental load scores (in terms of the NASA-TLX and SSSQ for the SWELL and WESAD datasets respectively) and based on the subtle changes in the participants' EDA and HRV signals. We also classified the stress based on the experiment conditions discussed in \cref{sec:dataset}. We trained and evaluated three stress prediction models:
\begin{enumerate}
	\setlength\itemsep{0em}
	\item  \textit{Person-specific models}\textemdash they were developed using Random Forest (RF) models (\cref{tab:rf-parameters}). All person-specific models were trained and tested exclusively on the physiological samples of the same person and validated using a 10-Folds cross-validation. 
	\item \textit{Generic models}\textemdash they were also developed using Random Forest (RF) models (\cref{tab:rf-parameters}). We used a
	Leave-One-Subject-Out Cross-Validation(LOSO-CV)to assess how a generic model would perform in predicting the stress of unseen people, (i.e., the people whose samples were not part of the training set) as follows: In a dataset of n subjects,  for each subject $S_i$, we trained the ML model on the data of \textit{(n-1)} subjects and validated its performance on the left-out subject $S_i$.    
	\item  \textit{Hybrid calibrated models}\textemdash as we expected (see discussion in \cref{sec:intro} on page~\pageref{sec:intro:underperformance} and \cref{sec:result}), the generic models performed poorly compared to the person-specific models. To mitigate this discrepancy, we devised a hybrid technique that derives a personalized stress prediction model from samples collected from a large population. The technique (\cref{algo:model-calibration}) consists of incorporating a few person-specific samples (the calibration samples) in a generic pool of physiological samples collected from a large group of people and to train a new model from this heterogeneous data. In this paper, for a dataset with N subjects, we used the calibration algorithm with $q=4$ and $n=N-q$, i.e., we reserved the physiological samples of four randomly selected subjects as \enquote{unseen subjects} and used data of the remaining $n=N-q$ subjects as the  \enquote{generic samples}. All calibration models were trained on a Extremely Randomized Trees models (ExtraTrees) whose key hyperparameters are summarized in \cref{tab:extra-trees-parameters}.
	
	\begin{algorithm}[htbp]
		\DontPrintSemicolon 
		\SetAlgoLined
		\KwIn{machine learning algorithm $h_m$}
		\KwData{
			\begin{itemize}
				\item Samples $sample_{generic}$ collected from $n$ persons 
				\item Calibration samples $sample_{calibration}$ that belong to $q$ unseen persons such that $q\ll n$
			\end{itemize}
		}
		\KwOut{trained calibrated model $h_m\prime$}
		\tcc{mix the calibration samples and the generic samples}
		$D^\prime \longleftarrow \emptyset$\;
		$D^\prime \longleftarrow shuffle(sample_{generic}\cup sample_{calibration})$\;
		\tcc{train the model  $h_m$ on dataset $D^\prime$}
		$h_m\prime \longleftarrow h_m(D^\prime)$\;
		\Return{$h_m\prime$}\;
		\caption{{\sc MODEL CALIBRATION} }
		\label{algo:model-calibration}
	\end{algorithm}
\end{enumerate}
We evaluated the classification models by computing their accuracy, precision, recall, and their  \textrm{$F_{1} score$} when tested on the test datasets. As for the regression models, their performance is evaluated by calculating their mean absolute error (MAE) and their root mean squared error (RMSE).

\section{RESULTS AND DISCUSSION} \label{sec:result}
\begin{figure*}[tbph]
	\centering
	\begin{subfigure}[b]{0.49\textwidth}
		\centering
		\includegraphics[width=1.0\linewidth]{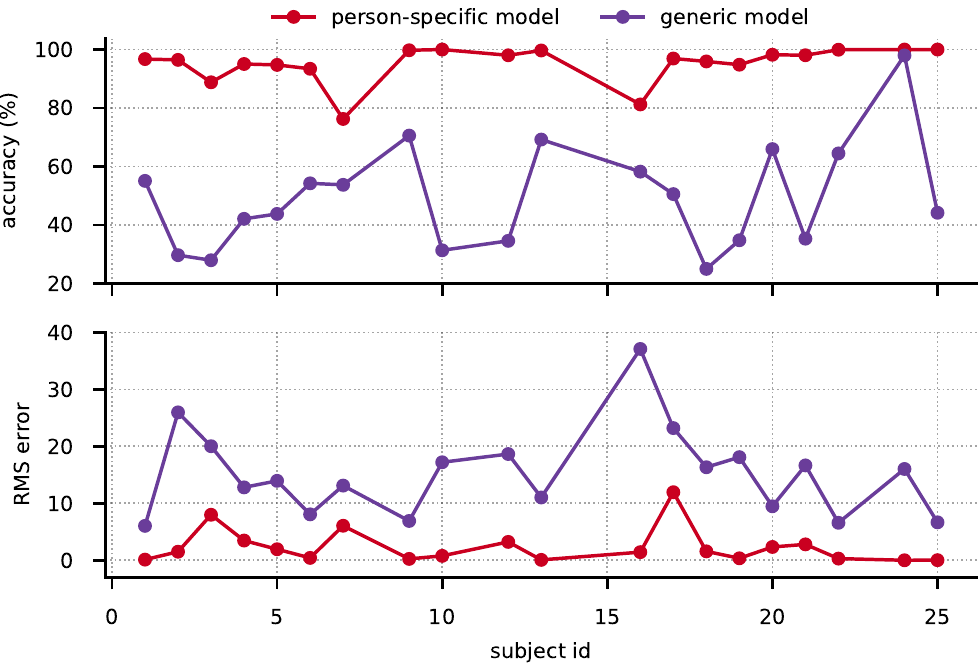}
		\caption{SWELL HRV dataset}
		\label{fig:swell-hrv-generic-vs-personal}
	\end{subfigure}
	\hfill
	\begin{subfigure}[b]{0.49\textwidth}
		\centering
		\includegraphics[width=0.9\linewidth]{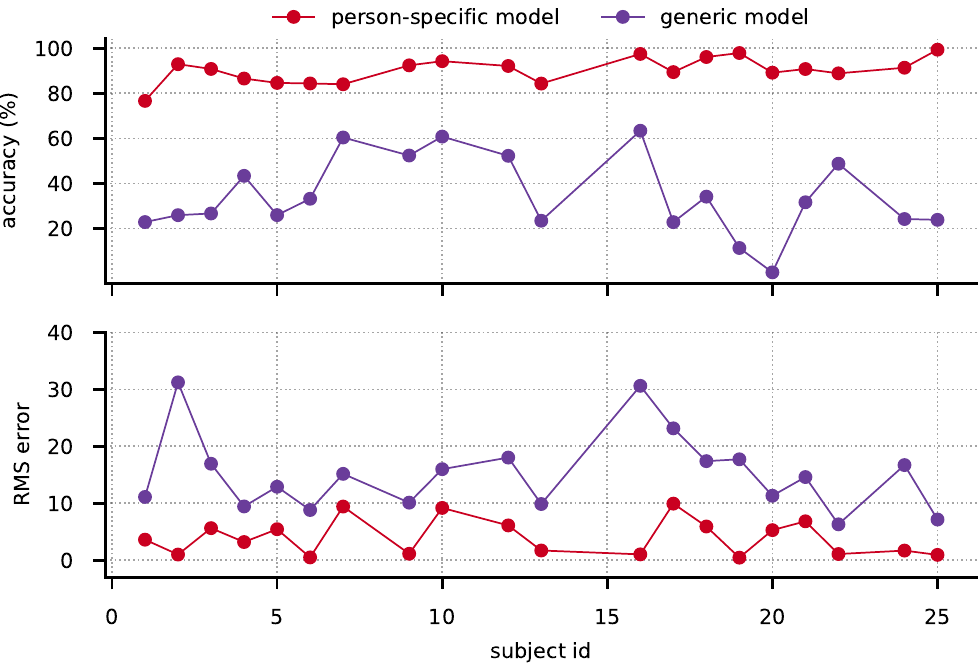}
		\caption{SWELL EDA dataset}
		\label{fig:swell-eda-generic-vs-personal}
	\end{subfigure}
	\caption{\textbf{Performance comparison between the person-specific and the generic models trained on the SWELL datasets}
		\newline For all subjects, the person-specific classification models (classification on three classes)achieved a high accuracy, and the regression models (based on NASA-TLX ($max=55.5, min=26.1, std=14.8$)) have a small a RMSE (e.g., $95.2\%\pm 0.5\%, 2.3\pm 0.1 RMSE$ for the HRV dataset). However, because of the inter-individual differences reacting to stress, all the generic models performed poorly (e.g., $42.5\%\pm 19.9\%, 15.3\pm 7.9 RMSE$ for the HRV signal), and there is a vast performance variation between the subjects.}
	\label{fig:swell-generic-vs-personal-performance}
\end{figure*}
\begin{figure*}
	\centering
	\begin{subfigure}[b]{0.49\textwidth}
		\centering
		\includegraphics[width=0.9\linewidth]{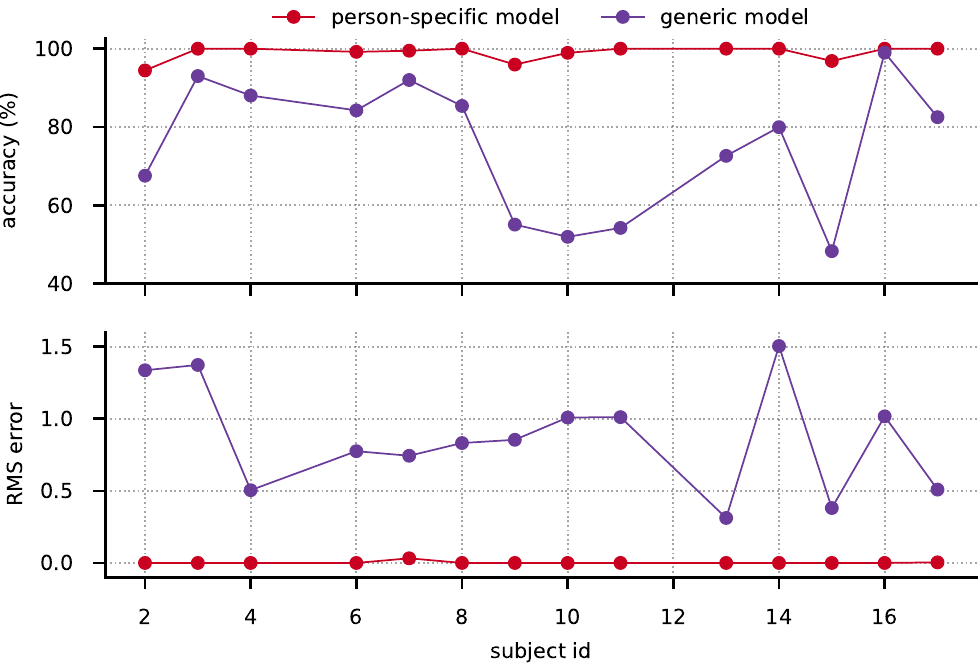}
		\caption{WESAD HRV dataset}
		\label{fig:wesad-hrv-generic-vs-personal}
	\end{subfigure}
	\hfill
	\begin{subfigure}[b]{0.49\textwidth}
		\centering
		\includegraphics[width=0.9\linewidth]{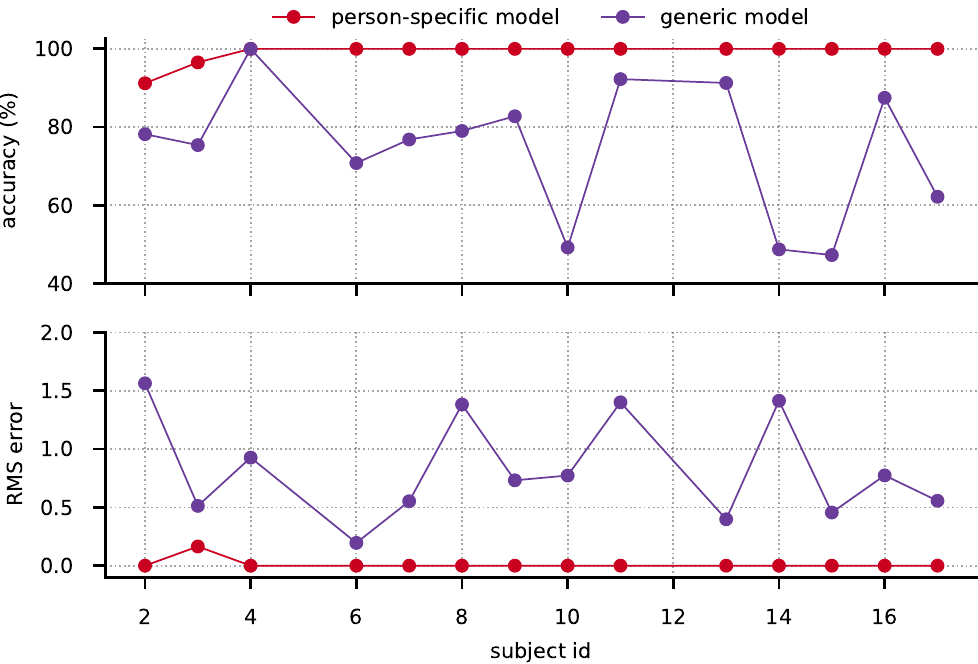}
		\caption{WESAD EDA dataset}
		\label{fig:wesad-generic-vs-personal-performance}
	\end{subfigure}
	\caption{\textbf{Performance comparison between the person-specific and the generic models trained on the WESAD datasets}
		\newline For all subjects, the person-specific classification models (classification on two classes) achieved a high accuracy, and the regression models (based on SSSQ ($max=3.9, min=3.0, std=0.8$)) have a low RMSE (e.g., $98.9\%\pm2.4\%, 0.002\pm 0.001 RMSE$ for the HRV signal). However, because of the differences in how different subjects react to stress, all the generic models performed poorly, and there is a vast performance variation between the subjects (e.g., $83.9\%\pm13.2\%, 0.8\pm 0.3 RMSE$ for the HRV signal).Also note that, compared to the SWELL datasets (\cref{fig:swell-generic-vs-personal-performance}), the classification models achieved seemingly a better performance because the dataset contains only two classes}
	\label{fig:wesad-generic-vs-personal-comparision}
	\vspace{-3mm}
\end{figure*}
\subsection{Individual differences in stress prediction} \label{sec:result-differences} 
All the person-specific models (i.e., the models that predict the stress of a preordained person) achieved an unrivaled performance. This high performance is, however, deceptive in that it would not generalize on yet unseen people. Indeed, the generic models (i.e., the models that predict the stress of any person) performed very poorly as shown in \cref{fig:swell-generic-vs-personal-performance,fig:wesad-generic-vs-personal-comparision}.
\par 
It is, of course, reasonable to assume the models over-fitted. However, there is no indication that this was the case. First, we validated all the person-specific models using a 10-fold cross-validation (CV) strategy, and it produced consistent predictions with a very low standard deviation between the 10-folds. K-Fold cross-validation provides an unbiased estimation of the performance of the model because it tests how well the k different parts of the training data perform on the model. Therefore, if there the models were over-fitted, the model would under-perform when tested on some folds. In our case, all folds achieved similar performance\myFooterTex{The interested readers are referred to the detailed tables in the supplementary material (see \cref{sec::supplement} for more details)}. Secondly, all the models use a very simple RF model (\cref{tab:rf-parameters}) that is less likely to overfit. We believe the models does not overfit because they consist of a large number of shallow trees (1000 trees, maximum depth=2) and that it has a small number of \textit{best split features}. A low \textit{best split features} allows the model to create more diverse and less correlated trees; therefore, the aggregation of the different trees results in a  model with a low generalization error variance and a high stability \cite{Probst2019}. Moreover, the trees are shallow (maximum depth=2) to reduce the model's complexity; thus, minimize overfitting. Finally, our results is similar to other published literature: in general, person-specific models achieve accuracy greater than 90\% \cite{Attaran2018},\cite{Nakashima2016} \cite{Liapis2015}\cite{Rigas2012} \cite{Melillo2011}\cite{Healey2005}\cite{Andre2008} while generic models always under-perform \cite{Nakashima2016}  \cite{Koldijk2018} \cite{Andre2008}. 
\par
\begin{figure*}[tbph!]
	\centering
	\begin{subfigure}[b]{0.48\textwidth}
		\centering
		
		\includegraphics[width=1.0\linewidth]{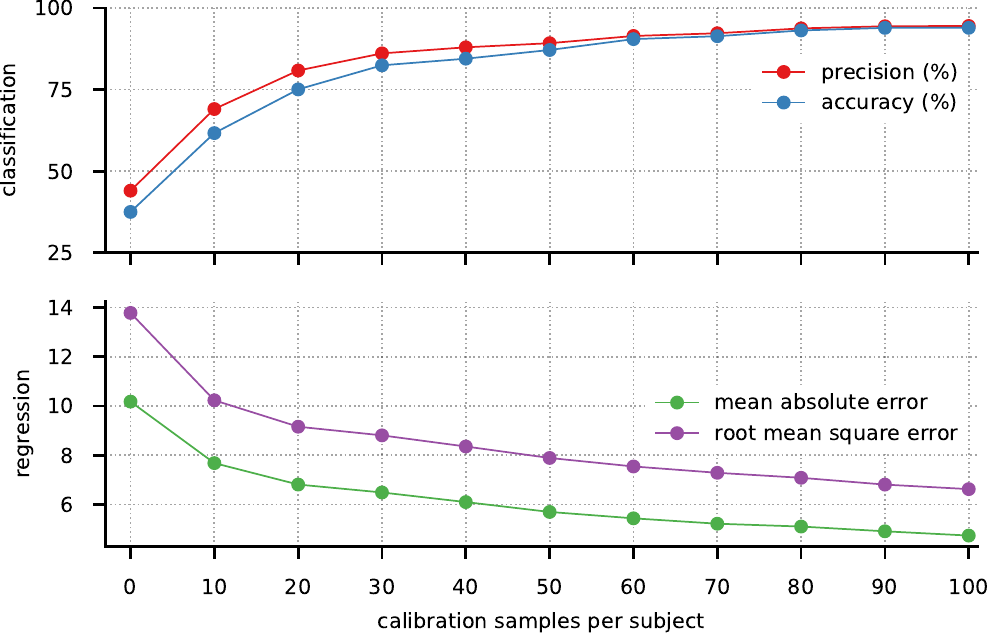}
		\caption{SWELL HRV dataset}
		\label{fig:swell-hrv-calibration}
	\end{subfigure}
	\hfill
	\begin{subfigure}[b]{0.48\textwidth}
		\centering
		\includegraphics[width=1.0\linewidth]{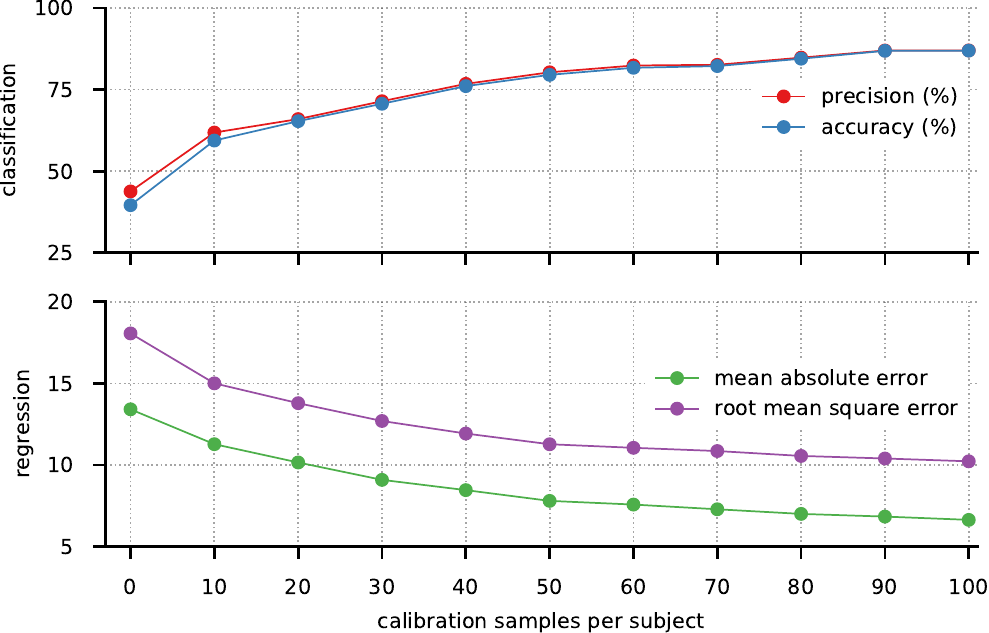}
		\caption{SWELL EDA dataset}
		\label{fig:swell-eda-calibration}
	\end{subfigure}
	\hfill
	\caption{\textbf{Performance of the hybrid model trained on the SWELL dataset} \newline
		without the calibration samples, both the regression and classification models performed crudely.  However, when a few person-specific calibration samples were used for calibration, their performance steadily improved}
	\label{fig:swell-hibrid-performance}
\end{figure*}
\begin{figure*}[tbph]
	\centering
	\begin{subfigure}[b]{0.48\textwidth}
		\centering
		
		\includegraphics[width=1.0\linewidth]{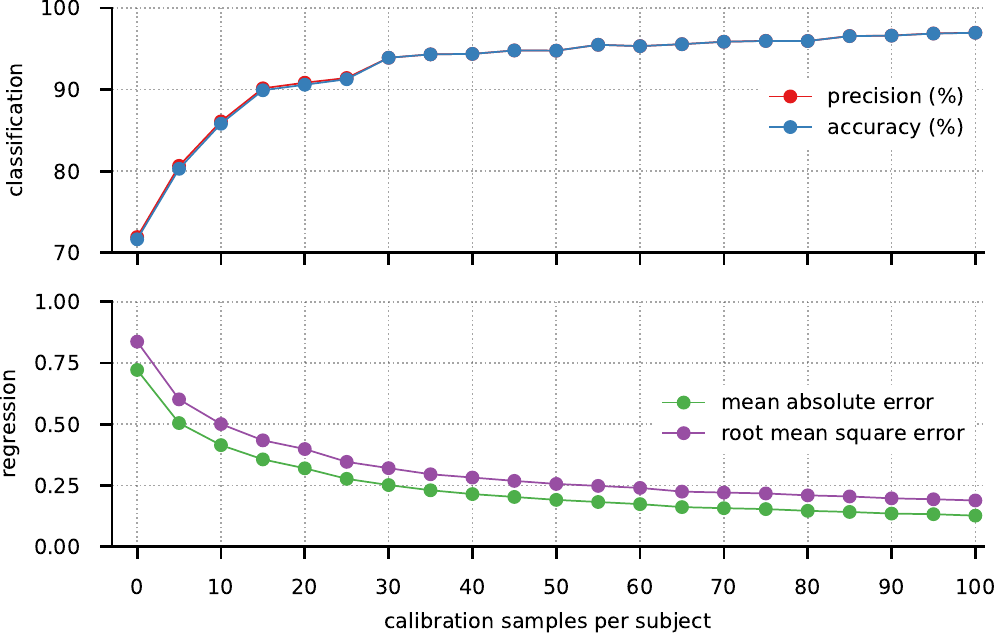}
		\caption{WESAD HRV dataset}
		\label{fig:wesad-hrv-calibration}
		
	\end{subfigure}
	\hfill
	\begin{subfigure}[b]{0.48\textwidth}
		\centering
		\includegraphics[width=1.0\linewidth]{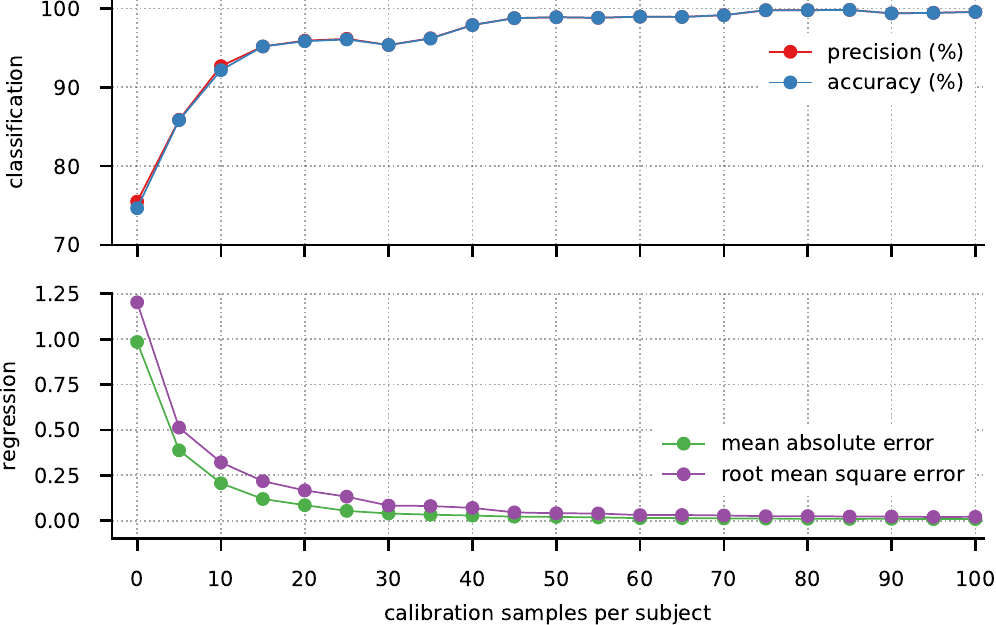}
		\caption{WESAD EDA dataset}
		\label{fig:wesad-eda-calibration}
	\end{subfigure}
	\caption{\textbf{Performance of the hybrid model trained on the WESAD dataset} \newline
		without the calibration samples, both the regression and classification models performed crudely.  However, when a few person-specific calibration samples were used for calibration, their performance steadily improved}
	\label{fig:wesad-hibrid-performance}
\end{figure*}
The drop in accuracy, when tested on unseen subjects, is also nothing out of the ordinary, as already explained (\cref{sec:intro} on page~\pageref{sec:intro:underperformance}). Indeed, the models cannot learn the inter-subject physiological differences in how people respond to the stressors. To double-check this verdict, we added a \textit{subject id} as a control prediction feature to the datasets. The \textit{subject id} was used to monitor the subject to whom each sample in the datasets belongs to and to probe how much each model is influenced by knowing the origin of each sample. The influence of the \textit{subject id} on the model is assessed by comparing the importance (in terms of a mean decrease in impurity (MDI)) of the  \textit{subject id} to that of other attributes of the dataset. The MDI score of an attribute reveals how much the said attribute contributes to making the final prediction of a model. We found that, in all datasets, the \textit{subject id} has the highest MDI; thus, is the most critical attribute for stress prediction. Additionally, as shown in \cref{fig:swell-generic-vs-personal-performance,fig:wesad-generic-vs-personal-comparision}, unlike the person-specific models, because each subject has a unique response to stress, the generic model's performance varies widely between the different subjects. Accordingly, using generic stress prediction models would lead to unpredictability and low performance compared to using person-specific models.
\par 
This discrepancy in performance highlights the far-reaching importance of inter-individual physiological differences that makes it hard for a generic stress prediction model to generalize to new unseen people. As already discussed by other researchers, one-size-fits-all stress prediction models cannot work well because people express stress differently. Furthermore, there is a wide gap in how the generic models performed on different subjects. This wide gap implies that, if a system uses a generic model for stress prediction, in practice, its prediction would seem virtually arbitrary and would make it very laborious to troubleshoot when the system has bugs. Therefore, an effective system would need to rely on non-economically viable person-specific models.

\subsection{Generic stress model calibration} \label{sec:result-calibration}
While it was possible to slightly increase the performance of the generic models (e.g., by using complex stacked models), it was clear that the performance of the person-specific models always dwarfs that of the person-independent models (\cref{fig:swell-generic-vs-personal-performance,fig:wesad-generic-vs-personal-comparision}). Furthermore, it was not possible to reliably use 
hyperparameters optimizations. The hyperparameters tuning is perverse guesswork and an erratic process given that the distribution of each subject is somehow unique; for that reason, finding hyperparameters for a model that work well for all subjects is a futile endeavor. 
\par 
In an attempt to improve the models' generalization on unseen people, we investigated how each model would perform if it knew little information about the previously unseen subjects. Consequently, we devised a technique that derive a personalized model from the data collected from a large group of people (see \cref{algo:model-calibration} on page~\pageref{algo:model-calibration}). In this paper, we used half of the data from $q=4$ randomly selected subjects as the \textit{calibration samples} and the remaining half is used to test the performance of the calibrated models. The data of the remaining $n=N-q$ subjects were used as the \textit{generic samples}. In one sense, the calibration samples serve as  \enquote{the fingerprints of a person}, i.e., they encode the \enquote{uniqueness} of an individual using tinny physiological samples of that person.
\par
When we applied this technique to stress prediction on the two datasets, the performance of all the models significantly increased, even when we only used a few calibration samples (see \cref{fig:swell-hibrid-performance,fig:wesad-hibrid-performance} for more details):
\begin{itemize}
	\item The root-mean-square error (RMSE) and mean absolute error (MAE) sharply dropped when we used a few calibration samples, and this is the case both for the model trained on the EDA datasets and the model trained on the HRV dataset. For instance, for the model trained on the HRV signal of the SWELL dataset, the MAE decreased from 10.1 to 7.6 when we only used 10 calibration samples per unseen subject. Likewise, this error dropped even more so when we used 100 calibration samples ($\text{mean absolute error =4.7, root-mean-square error=6.6}$). 
	\item In a like manner, the performance of the classification models noticeably increased when we used a few calibration samples. For instance, the model trained on the HRV signal of the SWELL dataset, the accuracy, the precision, and recall respectively increased from 37.5\%, 44.0\%,  and 37.50\% to 61.6\%, 69.0\% and 61.6\% when we used only 10 calibration samples per unseen subjects and culminated in a 93.9\% accuracy, 94.4\% precision and 93.9\% recall with100 calibration samples per subject. 
\end{itemize}
\par
The increase in performance due to the few person-specific calibration samples highlights the influence of the person-specific biometrics in predicting stress. In \cite{Lamichhane2017}, the authors showed that, when inter-individual physiological differences are not accounted for, a stress predictive model may perform no better than a model with no learning capability. Our result highly their findings. Nevertheless, all humans share a common hormonal response to stress \cite{Charmandari2005}, albeit a person's unique factors such as gender \cite{Wang2007a}, genetics\cite{Wust2004},  personality \cite{Childs2014}, weight \cite{Jayasinghe2014} and his coping ability \cite{Kogler2015} differentiate how each person reacts to stress. Previous researchers (e.g., \cite{Koldijk2018, Xu2015, Ramos2014}) have achieved notable improvements in generic stress prediction models by clustering the subjects based on their physiological or physical similarity. Their methods are, however, not practical for mass-product stress monitoring product because they rely on heuristic clustering methods, and there is no authoritative subject clustering criterion. Our proposed method is simpler and much cheaper for a real-world deployment (see discussion in \cref{sec:monitoring}) and performs much better than any previously proposed generic model improvement technique. 

\section{STRESS MONITORING IN OFFICES} \label{sec:monitoring}
\par
The above results suggest that, in order to design a real-world stress monitoring system, it would be beneficial to rethink the trade-off between spending effort on collecting data and training high performing, but costly person-specific model, versus using a hybrid model derived from a mixture of a few person-specific physiological samples with physiological samples collected from a large population. The latter approach is less expensive, more flexible for deployment, and delivers comparable performance to that of person-specific models.
\par 
The architecture and deployment of a stress monitoring system that uses this technique will undoubtedly involve a lot of technical challenges that are beyond the scope of this paper. We encourage the interested reader to examine \cite{Can2019a}\cite{Alberdi2016} for an exhaustive overview of these challenges.  One of the biggest challenges is perhaps how to collect the required physiological signals unobtrusively. Indeed, the system should not interfere with a person's routine. At the same time, it should record the physiological signals meticulously, accurately and at an adequate sampling frequency because the quality of the physiological data affects the performance of the stress prediction models \cite{Chowdhury2018}. These stringent requirements necessitate making conflicting compromises. For instance, while an HRV signal recorded using the chest leads is always of the highest quality, its recording would hinder the person's normal life. Alternatively, the HRV signal could be obtained using a lower quality but less invasive PPG signal recorded from the person's wrist. There exist many wearable devices (e.g., smart-watches and fitness trackers) with built-in PPG sensors. For example, the Empatica E4 wristband\footnote{\url{https://www.empatica.com/research/e4/}} might serve for this purpose. The device boasts of a high-resolution EDA sensor with a strong steel electrode that can continuously record both the tonic and phasic changes in the skin conductance. As discussed in a recent article \cite{Menghini2019}, the Empatica E4 wrist band has an adequate accuracy in recording HRV in seated rest, paced breathing, and recovery conditions. However, it is not very reliable when its wearer makes wrist movements. 
\par 
\begin{figure*}[htbp]
	\centering
	\includegraphics[width=1.0\linewidth]{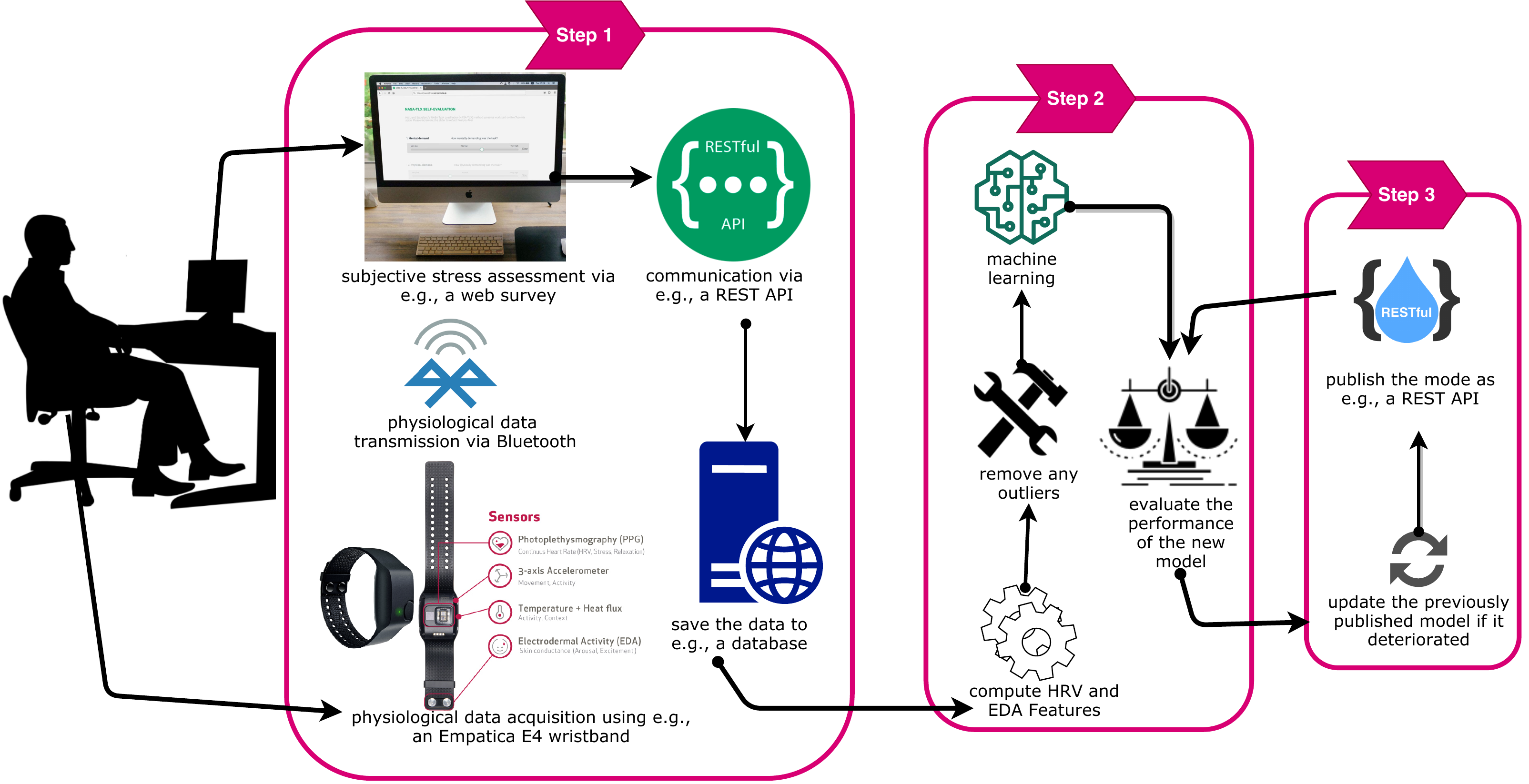}
	\caption{\textbf{A simplified pipeline for a continuous stress monitoring model} \newline A person’s photoplethysmogram (PPG) and EDA signals are recorded using a wristband device. The signals are sent to a computing device where appropriate features  (e.g., \cref{tab:hrv-eda-features} on page~\pageref{tab:hrv-eda-features}) are computed, preprocessed (e.g., data cleaning, and rebalancing) and sent to a remote server where they are used to predict the person’s stress. For calibration purses, the person also periodically provide self-assessment of his stress (e.g., via a web survey after the completion of his work). This feedback is used to train a personalized stress prediction model, which is published and consumed as a RESTful API. When the model deteriorates,  it is automatically updated based on the periodic self-evaluations the system received from its users.}
	\label{fig:incrementalstressprediction}
\end{figure*}
Another challenge is how to deploy the stress prediction models. The recent reviews on stress recognition\cite{Carneiro2019}\cite{Schmidt2018} unanimously concluded that due to the physiological difference in how people react to stress, a stress monitoring system should adapt to every individual's physiological needs. The simple, and likely, the most accurate approach is to deploy each person's stress prediction model as a web service (e.g., Representational State Transfer (REST) web service) that can be consumed to predict the person's stress. Regrettably, such an approach is daunting, time-consuming, and expensive because, in.e.g., office environment, it would require to collect, clean and label new data and train a new model of each office employee. Moreover, once deployed, the resulting stress monitoring system will unquestionably not perform as expected because its performance would deteriorate with time considering that a person's stress is dynamic and affected by many factors \cite{Schneiderman2008, Johnson1992}. Consequently, with this approach, a real-world system will need to periodically start over and collect, label, and train new models for each user to prevent the system from the anticipated performance degradation.
\par 
As implied by the results of this paper (see \cref{sec:result-calibration}), an alternative and cost-effective method would be to derive a high performing model from a combination of generic samples collected from a large population and few person-specific calibration samples. It would also be beneficial to automate this process entirely. As an illustration, after training and testing a generic stress prediction model, it could be possible to create an automatic self-updating stress prediction model pipeline, depicted in \cref{fig:incrementalstressprediction}, as follows:

\begin{enumerate}[leftmargin=1.2cm, label=\textbf{STEP \Roman*}]
	\setlength\itemsep{0em}
	\item \textit{calibration samples collection} \textemdash  Once the stress monitoring system is deployed, at the beginning (at this point, it uses only a generic model), it is primordial that its users take several self-evaluation surveys in different working conditions to allow the collection of self-evaluation ground-truths that reflect the broader rangers of stressors that its users will likely go through. At the same time, each user's physiological signals are recorded using an unobtrusive wearable device (e.g., an Empatica E4 wristband) and saved in a database. Once the system has collected enough calibration samples from the users, it would automatically create each user's personalized model by training a new model on a combination of the new user-specific data with the data that was used to train the generic model as shown in \cref{algo:model-calibration}.
	\item \textit{continuous machine learning}\label{step-cont-learning} \textemdash After these personalized models are deployed, the system would periodically remind its users to provide additional calibration samples by taking shorts self-report survey periodically (e.g., via a web survey every time he/she finished a task) to give more feedback data to improve the user's personalized model. Indeed, with time, the models will be prone to the effect of concept drift \cite{Gama2014}, i.e., they will become stale because their input data unpredictably change over time. In stress prediction, model drifting is particularly inevitable because stress is inherently dynamic \cite{Johnson1992}. The models, thus need to adapt to the new changes. For example, when the system has received a specific number of new calibration samples from a user, it would automatically test their accuracy against the existing model. If this prediction indicates a deterioration of the model, the system will need to update the model to reverse the drift. There are many ways to achieve this. One approach would be to train a new model on a combination of the data of the generic model and the new calibration samples. This approach would be, however, computationally expensive and require significant time to retrain each user's model. Depending on the system, it would be instead more appropriate to incrementally train the existing model as the new data is received \cite{Gepperth2016}. This approach is faster because it does not require retraining the whole model when new data come in. Instead, it extends the existing model by, e.g., combining the new data with a subset of the old data \cite{Castro2018}. Nevertheless, It is important to note that many machine learning algorithms do not support incremental learning and that, unless there is rigorous monitoring of the system, incremental learning may introduce nefarious predicaments \cite{Gepperth2016}.
	\item \textit{calibrated model deployment} \textemdash the model is published as, e.g., a REST Application Program Interface (REST API) and periodically updated depending on its performance as discussed in \ref{step-cont-learning} above.
\end{enumerate}
\par 
Although there is a need to validate our assumptions, we believe that developing a continuous stress monitoring system based on this strategy would present the following benefits over existing approaches: 
\begin{itemize}
	\item \textit{lower cost} \textemdash for practicality, the existing approaches would require collecting and labeling the training data for each user. This process is costly and would require expensive installation, support, and maintenance services costs. Our approach would likely be less expensive because there will be no need to collect large quantities of new data from each user. Instead, only a few user-specific samples are required. 
	\item\textit{practicality} \textemdash all high accuracy stress prediction methods rely on person-specific models. As already discussed, this approach is suboptimal when applied to new unseen people. The alternative is to create person-specific models. While this approach performs excellently in predicting stress, it is not practical in real-world settings because it is not scalable to many users, would be very costly to implement, and, most importantly, this approach is rigid and not flexible to the expected dynamic changes in each user's stress. The proposed approach achieves a stress prediction accuracy that is comparable to that achieved by subject-dependent models and yet, presents enticing large scale deployment benefits. 
	\item \textit{straightforward deployment} \textemdash once deployed, each user's person-specific model can be generated using negligible user-specific samples that can be unobtrusively collected using, e.g., an approach proposed in \cite{Bush2014} in which each user can self-evaluate (in terms of NASA-TLX and SSSQ) his stress level via a smartphone application. The self-evaluation would serve as a person-specific calibration to the generic model. Over time, when the model degrades due to the person's dynamics in stress, a few new physiological samples would be collected and used to train and update each person's model periodically. 
\end{itemize}
\par
Although the results of this study are encouraging, there are still many limitations. Notably, the study did not validate the proposed approach in real-world settings, and it reached its conclusion using only two datasets with a small homogeneous group of subjects. Further, designing a continuous stress monitoring system using the proposed approach requires extraordinary care because external factors can influence both the EDA and the HRV. In particular, the EDA signal, while it is often heralded as one of the best indicators of stress \cite{Alberdi2016, Setz2010}, it has significant drawbacks. The EDA is a result of electrical changes that happen when the skin receives signals from the nervous system. Under stress, the skin's conductance changes due to a subtle increase in sweat that lead to a decrease in the skin's electrical resistance.  The variation in skin conductivity is, however, influenced by other unrelated factors such as the person's hydration, the ambient temperature, and the ambient humidity. Moreover, for the same person, an EDA signal may fluctuate from one day to another \cite{Bakker2011}. Additionally, because stress is intrinsically multifaceted (it consists of physiological, behavioral and affective response), as highlighted in \cite{Panagakis2018}, it is imperative to take into consideration its context (i.e., where, what, when, who, why, and how). This approach, as shown in \cite{Gjoreski2017}, may yield better and predictable results even when tested in real-life conditions. 
\par
It is also important to highlight that the deployment of a stress monitoring system based on our approach still poses technical and cost challenges. The system would require considerable upfront investments and would be undoubtedly out of a budget of a small business. However, the investment might be well worth it for a large business. In our previous studies, we showed that it is possible to predict people's thermal comfort using the variations in their HRV \cite{Nkurikiyeyezu2019a}\cite{Nkurikiyeyezu2018b} \cite{Nkurikiyeyezu2017a}, and highlighted the energy-saving potential of this approach \cite{Nkurikiyeyezu2018a}. Therefore, the positive spillovers that might result in using the system may outstrip the initial investment because, in a responsive smart office, the system can be used as part a of a  multipurpose system that uses the office occupants' physiological signals for preventive medicine, stress management, and provides an efficient thermal comfort at low energy. Additionally, there exist enabling technologies that would make these challenges a little bit easier. For example, IBM's Watson Studio\footnote{\url{https://www.ibm.com/cloud/machine-learning}} offers tools that simplify developing and deploying predictive models. In our proposed stress monitoring system, Watson Studio could be used \textemdash and requires little or no programming experience \textemdash to automate steps 1 and 2 (see \cref{fig:incrementalstressprediction} ) including model deterioration monitoring and deployment as a REST API. 
\section{CONCLUSION}
Despite an extensive body of literature on stress recognition, and notwithstanding the potential economic and health benefit of stress monitoring, there exists no robust real-world stress recognition system. The most reliable and uncompromising methods use a fusion of multi-modal signals (e.g., physiological (such as HRV, EDA, EEG, EMG, skin temperature, respiration, pupil diameters, eye gaze), behavioral (keystrokes and mouse dynamics, and sitting posture), facial expression, speech patterns, and mobile phone use patterns). This approach, however, raises both practical challenges (e.g., real-time multi-modal data acquisition, data fusion, and data integration) and user privacy concerns (e.g., the implication of recording a person's computer keystrokes, his video and his speech), and,  are not feasible in the real-world settings because of company-wide computer security policies or due to international workplace privacy laws. 
\par 
On the contrary, the most practical stress monitoring methods that use physiological signals are idiosyncratic because stress is inherently subjective and is felt differently depending on the person. Therefore, methods that use ML model that uses physiological signals fail to generalize well when predicting the stress of new unseen people. Thus, they are not suitable for a real-world stress monitoring system. Only person-specific models are accurate enough for this task. Unfortunately, unlike the generic models, person-specific models are inflexible and costly to deploy in real-world settings because they require collecting new data and training a new model for every user of the system. In an office environment, this entails spending precious resources to collect and train a new model for every employee. Moreover, because stress is inherently dynamic, these models will need expensive periodic updates to collect and retrain every model to prevent the system from deterioration due to concept drift.
\par 
In this paper, we proposed a cost-effective hybrid stress prediction approach. Our method takes its foundation on the fact that humans share similar hormonal responses to stress. However, every person possesses unique factors (e.g., gender, age, weight, and copying ability) that differentiate the person from others. Therefore, we hypothesized that it could be possible to improve the generalization performance of a generic stress prediction model trained on a large population by deriving a personalized model from a combination of samples collected from a  large group of people with a few person-specific samples. In a sense, the calibration samples serve as the \say{fingerprint} of a person and they introduce his/her \say{uniqueness} into the new model. 
\par 
We tested our method on two stress datasets and found that our approach performed much better than the generic models. Furthermore, we surmised that, in order to create a practical stress monitoring system, this approach would be cost-effective and practical to deploy in real-world settings and discussed some of its technical limitations.

	\section*{SUPPLEMENTARY MATERIAL} \label{sec::supplement}
	\noindent Additional  supporting  information  are available online\myFooterTex{Available at \url{\datasetUrl}} in our public repository. The repository contains more detailed information and the source code to replicate our finding:
	\begin{itemize}
		\item Source code we developed for this research
		\item Dataset of the computed HRV and EDA features
		\item HRV and EDA feature importance with and without the \textit{subject\_id} added to the datasets (see \cref{sec:feature-engineering})
		\item Tables of the performance of the person-specific and generic models models (refer to \cref{sec:result-differences})
		\item Tables of the of performance of the calibrated models (see details in  \cref{sec:result-calibration})
	\end{itemize}
{\balance
	\microtypesetup{protrusion=false} 
	\bibliographystyle{IEEEtran}
	\bibliography{tac2019.bib}
	\microtypesetup{protrusion=true} 
}
\end{document}